\documentclass[sigconf]{acmart}

\AtBeginDocument{%
  \providecommand\BibTeX{{%
    \normalfont B\kern-0.5em{\scshape i\kern-0.25em b}\kern-0.8em\TeX}}}

\setcopyright{acmlicensed}
\copyrightyear{2025}
\acmYear{2025}
\setcopyright{acmlicensed}
\acmConference[KDD '25]{Proceedings of the 31st ACM SIGKDD Conference on Knowledge Discovery and Data Mining V.2}{August 3--7, 2025}{Toronto, ON, Canada}
\acmBooktitle{Proceedings of the 31st ACM SIGKDD Conference on Knowledge Discovery and Data Mining V.2 (KDD '25), August 3--7, 2025, Toronto, ON, Canada}
\acmISBN{979-8-4007-1454-2/25/08}
\acmDOI{10.1145/3711896.3737219}

\acmSubmissionID{247}

\usepackage{latexsym}

\usepackage{microtype}
\usepackage{balance}

\usepackage[utf8]{inputenc} 
\usepackage[OT1]{fontenc} 
\usepackage{hyperref} 
\usepackage{url} 
\usepackage{booktabs} 
\usepackage{amsfonts} 
\usepackage{nicefrac} 
\usepackage{microtype} 
\usepackage{xcolor} 
\usepackage{CJKutf8}
\usepackage{multirow}
\usepackage{amsmath}
\usepackage{amsthm}
\usepackage{bbding}
\usepackage{graphicx}
\usepackage{makecell}

\begin{document}

\title[FinBERT2: A Specialized Bidirectional Encoder for Bridging the Gap in Finance-Specific \\ Deployment of Large Language Models]{FinBERT2: A Specialized Bidirectional Encoder for Bridging the Gap in Finance-Specific Deployment of Large Language Models}


\author{Xuan Xu}
\affiliation{%
  \institution{Beijing University of Posts and Telecommunications}
  \city{Beijing}
  \country{China}}
\email{sh22xuxuan@bupt.edu.cn}
\authornote{XX, FW contributed equally to this research.}

\author{Fufang Wen}
\affiliation{%
  \institution{Beijing Value Simplex Technology Co.Ltd. }
  \city{Beijing}
  \country{China}}
\email{wenfufang@entropyreduce.com}
\authornotemark[1]

\author{Beilin Chu}
\affiliation{%
  \institution{Beijing University of Posts and Telecommunications}
  \city{Beijing}
  \country{China}}
  \email{beilin.chu@bupt.edu.cn}
  
\author{Zhibing Fu}
\affiliation{%
  \institution{Beijing Value Simplex Technology Co.Ltd. }
  \city{Beijing}
  \country{China}}
\email{fuzhibing@entropyreduce.com}

\author{Qinhong Lin}
\affiliation{%
  \institution{Beijing University of Posts and Telecommunications}
  \city{Beijing}
  \country{China}}
\email{greenred99@bupt.edu.cn}

\author{Jiaqi Liu}
\affiliation{%
  \institution{Beijing Value Simplex Technology Co.Ltd. }
  \city{Beijing}
  \country{China}}
\email{liujiaqi@entropyreduce.com}

\author{Binjie Fei}
\affiliation{%
  \institution{Beijing Value Simplex Technology Co.Ltd. }
  \city{Beijing}
  \country{China}}
\email{feibj@entropyreduce.com}

\author{Yu Li}
\affiliation{%
  \institution{Beijing Value Simplex Technology Co.Ltd. }
  \city{Beijing}
  \country{China}}
\email{liyu@entropyreduce.com}
\authornotemark[2]

\author{Linna Zhou}
\affiliation{%
  \institution{Beijing University of Posts and Telecommunications}
  \city{Beijing}
  \country{China}}
\email{zhoulinna@bupt.edu.cn}
\authornotemark[2]

\author{Zhongliang Yang}
\affiliation{%
  \institution{Beijing University of Posts and Telecommunications}
  \city{Beijing}
  \country{China}}
\email{yangzl@bupt.edu.cn}
\authornote{Corresponding Authors: YL, LZ and ZY}

\renewcommand{\shortauthors}{Xuan Xu, et al.}

\begin{abstract}
  In natural language processing (NLP), the focus has shifted from encoder-only tiny language models like BERT to decoder-only large language models(LLMs) such as GPT-3. However, LLMs' practical application in the financial sector has revealed three limitations: (1) LLMs often perform worse than fine-tuned BERT on discriminative tasks despite costing much higher computational resources, such as market sentiment analysis in financial reports; (2) Application on generative tasks heavily relies on retrieval augmented generation (RAG) methods to provide current and specialized information, with general retrievers showing suboptimal performance on domain-specific retrieval tasks; (3) There are additional inadequacies in other feature-based scenarios, such as topic modeling. We introduce FinBERT2, a specialized bidirectional encoder pretrained on a high-quality, financial-specific corpus of 32b tokens. This represents the largest known Chinese financial pretraining corpus for models of this parameter size. As a better backbone, FinBERT2 can bridge the gap in the financial-specific deployment of LLMs through the following achievements: (1) Discriminative fine-tuned models (Fin-Labelers) outperform other (Fin)BERT variants by 0.4\%-3.3\% and leading LLMs by 9.7\%-12.3\% on average across five financial classification tasks. (2) Contrastive fine-tuned models (Fin-Retrievers) outperform both open-source (e.g., +6.8\% avg improvement over BGE-base-zh) and proprietary (e.g., +4.2\% avg improvement over OpenAI's text-embedding-3-large) embedders across five financial retrieval tasks; (3) Building on FinBERT2 variants, we construct the Fin-TopicModel, which enables superior clustering and topic representation for financial titles. Our work revisits financial BERT models through comparative analysis with contemporary LLMs and offers practical insights for effectively utilizing FinBERT in the LLMs era.\footnote{Our work will be accessible at \url{https://github.com/Value Simplex/FinBERT2}}

\end{abstract}



\maketitle

\section{Introduction}

Early LLMs primarily relied on encoder-only architectures with masked language modeling (MLM), such as BERT \citep{devlinBERTPretrainingDeep2019}, RoBERTa \citep{liuRoBERTaRobustlyOptimized2019}, and XLM \citep{lample2019cross}. However, between 2018 and 2021, the field shifted from single-task fine-tuning to large-scale multi-task learning. GPT-3 \citep{bommasani2021opportunities} demonstrated that scaling \citep{zhang2022examining} could significantly narrow the performance gap between autoregressive and other architectures. Moreover, autoregressive models offer advantages such as greater task adaptability, a unified modeling paradigm \citep{wang2022language}, and reduced engineering complexity. Consequently, decoder-only models have become the dominant paradigm in the development of LLMs.
 
Similarly, the financial domain has witnessed a shift from early FinBERT \citep{araciFinBERTFinancialSentiment2019,desolaFinBERTPretrainedModel2019,huangFinBERTLargeLanguage2023} to large-scale FinLLMs ranging from billions to trillions of parameters, such as FinGPT \citep{wangFinGPTInstructionTuning2023}, BloombergGPT \citep{wuBloombergGPTLargeLanguage2023}, and FinLlama \citep{konstantinidisFinLlamaFinancialSentiment2024}. These models leverage domain-specific data and undergo post-training on financial text corpora, enhancing their understanding and generation capabilities in financial applications. However, LLMs, including fin-adapted versions, do not fully replace BERT. They still face limitations in real-world deployment.\footnote{Classical BERT was once considered an LLM, but BERT less than 1 billion parameters is now seen as a Tiny LM and no longer meets the criteria of an LLM}

\begin{figure*}[htbp]
    \centering
         \includegraphics[
        width=\textwidth,
        trim={50pt 560pt 30pt 60pt},  
        clip
    ]{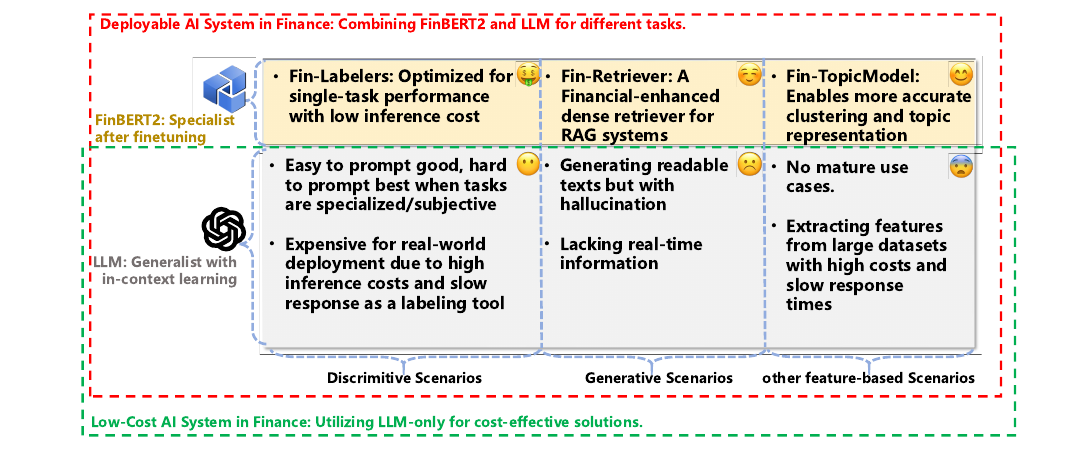}
    \caption{Illustration of how FinBERT2 bridges the gap in finance-specific deployment. The comparison presents two solution: (1) A deployable AI system combining FinBERT2 (serve as domain specialist after fine-tuning) with LLMs (serve as general-purpose model with in-context learning capabilities), and (2) A conventional (Fin)LLM-only system.
}
\end{figure*}

Firstly, while LLMs exhibit strong generalization and robustness as large-scale multi-task models, they are not always optimal for specific NLU tasks. In certain single-task scenarios, fine-tuned BERT-base models often outperform them. For instance, \citet{koconChatGPTJackAll2023}
evaluated ChatGPT on 25 analytical NLP tasks and found that, compared to state-of-the-art (SOTA) methods, its zero-shot and few-shot performance dropped by approximately 25\% on average. The decline was even more pronounced for complex tasks. Similarly, \citet{huBadActorGood2024} reported similarly in fake news detection, GPT-3.5 underperformed compared to specialized, smaller models like BERT.
Moreover, LLMs are costly and slow for data-intensive tasks, such as labeling financial reports, while smaller BERT-like models (0.1B parameters) are more efficient.

Secondly, in generative scenarios requiring external financial knowledge, such as real-time question answering (QA) on financial reports, LLMs rely on embedding-based retrieval \citep{huang2020embedding} to ensure accuracy and timeliness. This approach, known as retrieval-augmented generation (RAG) \citep{lewis2020retrieval}, necessitates efficient offline retrieval. Consequently, dense retrievers (DRs) built on dual-encoder BERT architectures, such as M3E \citep{Wang2023M3E}, BGE \citep{xiaoCPackPackagedResources2023}, and BCE \citep{youdao_bcembedding_2023}, have become mainstream.
These DRs achieve high retrieval accuracy on general benchmarks \citep{muennighoffMTEBMassiveText2023} through in-batch negative learning on large-scale weakly supervised sentence pairs and contrastive fine-tuning with mined hard negatives. However, despite extensive training, they often underperform in specialized domains such as finance \citep{tang2025needdomainspecificembeddingmodels}.

Thirdly, beyond retrieval, LLMs are less effective \citep{qorib-etal-2024-decoder} or not mature in feature-based tasks such as clustering-based topic modeling \citep{angelov2020top2vec}, measuring earnings surprises and market reactions \citep{meursault2023pead}, and extracting factors of stock price prediction \citep{10.1145/3677052.3698618}. Generative models often lack mature use cases for these applications, as they require compact and efficient feature encoding \citep{search2019bert} and flexible fine-tuning for task-specific adaptations. For instance, topic modeling prioritizes industry-specific features, while stock return prediction relies on sentiment features. However, decoder-only LLMs struggle to meet these requirements due to their inherent architectural constraints.

The above challenges hinder the applications in small and medium-sized financial enterprises, where scenarios are often specialized and diverse. To address this, we re-evaluate the value of lightweight, localized, and customizable FinBERT models and propose a hybrid architecture that integrates FinBERTs (as domain experts mitigating these limitations) with LLMs (as general-purpose generative models with in-context learning capabilities).
Specifically, we pre-train FinBERT2, an enhanced version of its predecessor FinBERT1 \citep{FinBERT}. FinBERT2 is trained on a carefully curated Chinese financial corpus comprising 32B tokens and further optimized for downstream tasks such as labeling, retrieval, and topic representation. In these tasks, FinBERT2 can effectively replace, assist, or complement LLMs, respectively, offering a more efficient and deployable NLP system for financial applications. Our contributions can be summarized as follows:

1) We pre-trained FinBERT2 on a 32B token Chinese financial corpus to inject domain knowledge. To the best of our knowledge, this is the largest pre-training corpus for a BERT-like language model in the Chinese finance domain, and it is trained with a finance-customized tokenizer.

2) As a more efficient and high-performance alternative to labeling, Fin-Labelers outperform other (Fin)BERT variants by 0.4\%–3.3\% and leading LLMs (e.g., GPT-4-turbo, Claude 3.5 Sonnet, Qwen2) by 9.7\%–12.3\% on average across five financial classification tasks.\footnote{The term "Labeler" highlights its role in the LLM era—fine-tuning BERT for task-specific, large-scale, real-time text processing.}

3) As an enhanced RAG assistant, Fin-Retrievers surpass both open-source and proprietary embedding models. They achieve an average improvement of +6.8\% over BGE-base-zh and +4.2\% over OpenAI’s text-embedding-3-large across five financial retrieval tasks.

4) In feature-based applications such as topic modeling, Fin-TopicModel, built on FinBERT2 variants, enables superior clustering and topic representation for financial titles.

\section{Related Works}

\subsection{FinLMs}

Since 2019, specialized models have emerged to tackle the complexities of financial text. FinBERTs (\citet{araciFinBERTFinancialSentiment2019}, \citet{desolaFinBERTPretrainedModel2019}, and \citet{huangFinBERTLargeLanguage2023}) have demonstrated proficiency respectively in financial sentiment analysis, document processing, and information extraction from financial texts. In the Chinese community, \citet{FinBERT}'s FinBERT was open-sourced, and its F1 scores in different financial tasks were significantly improved compared to the vanilla BERT. Another influential work, Mengzi-fin (\citet{zhangMengziLightweightIngenious2021}) further enriched the field of Chinese FinBERTs.
In parallel, larger LMs have ventured into the financial domain, tailoring their capabilities to specific financial tasks.
BloombergGPT (\citet{wuBloombergGPTLargeLanguage2023}) is characterized by a 50B-parameter model trained on a 363B token financial corpus.
The FinTral suite by \citet{bhatiaFinTralFamilyGPT42024}, based on Mistral-7b and multimodal data, outperformed GPT-4 on several tasks including Sentiment Analysis (SA), Named Entity Recognition (NER) etc. FinLlama (\citet{konstantinidisFinLlamaFinancialSentiment2024}), derived from Llama-2-7b, improved sentiment classification accuracy and quantified sentiment strength.
Both FinBERTs and FinLLMs highlight the advantages of in-domain pre-training. In these works, although FinLLMs have greater potential and application scenarios, relatively small specialized models trained on more financial corpus are more parameter-efficient, either matching or outperforming much larger language models.

\subsection{Dense Retrievers}
Dense retrievers (DRs), also called text embedders, typically use a BERT-based dual-encoder architecture that learns by minimizing the distance between positive sample pairs and maximizing the distance between negative sample pairs. This independence of document representation from query representation accommodates the offline computational demands of large corpora. Pre-training tasks designed specifically for retrieval (passage ranking), such as ICT \citep{lee2019latent}, BFS \citep{chang2020pre}, and RetroMAE \citep{xiaoRetroMAEPreTrainingRetrievaloriented2022} have proven effective, but it has also been shown that carefully fine-tuning a vanilla BERT model can also outperform these approaches \citep{karpukhin2020dense,ma2024drop} 
Recent studies have explored the construction of generalized text representation models through large-scale contrastive learning \citep{neelakantan2022text,suOneEmbedderAny2023,wangTextEmbeddingsWeaklySupervised2024}. These works mostly follow a multi-stage training approach: i.e., pre-training on large-scale weakly supervised text pairs in an in-batch-negative manner, and supervised fine-tuning on triplets with hard negatives to further fit to popular benchmarks \citep{thakur2021beir,muennighoffMTEBMassiveText2023,xiaoCPackPackagedResources2023}. Despite extensive training, out-of-domain generalization remains limited. These DRs even fail to reach the BM25 level without further fine-tuning on labeled datasets.

\begin{figure*}[htbp]
    \centering
         \includegraphics[
        width=\textwidth,
        trim={110pt 260pt 110pt 140pt},  
        clip
    ]{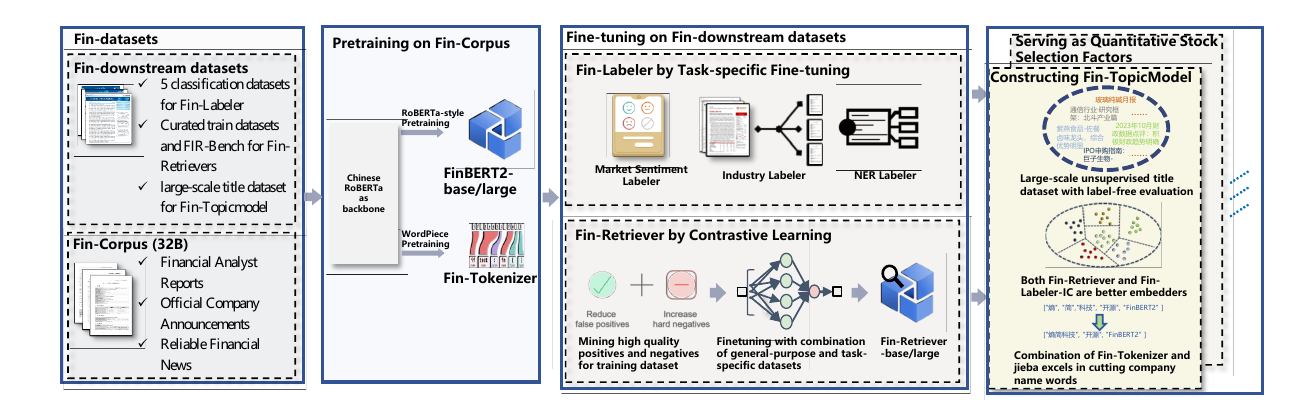}
    \caption{A overview of our FinBERT2 work.
}
\label{fig:overview}
\end{figure*}

\section{Methods}
\subsection{Overview of FinBERT2}
As shown in \autoref{fig:overview}, our work starts from the data layer consisting of Fin-Corpus and Fin-downstream datasets, through the foundation layer where produces Fin-Tokenizer and FinBERT2-base/large, to the downstream application layer featuring three main components: (1) Fin-Labeler, fine-tuned for five downstream financial tasks including market sentiment classification, industry classification, and named entity recognition (NER);
(2) Fin-Retriever, trained via contrastive learning for five financial retrieval tasks; (3) Fin-TopicModel, an enhanced version of TopicModel, integrating key components and improvements derived from other FinBERT2 variants.

\subsection{Pre-training of FinBERT2} 
\subsubsection{Fin-Corpus for Pre-training}
The FinBERT2 model has seen a significant expansion in its pre-training corpus compared to FinBERT1 \citep{FinBERT}, with the total token count increasing to 32B (99 G).

\textbf{1) Analyst Reports Corpus (16B tokens, 53G)}: We have compiled a collection of 2.6 million financial analysis reports, encompassing over twenty types of reports such as stock/futures research, industry analysis and institutional commentaries. The dataset spans the past 15 years and has undergone detailed data cleansing.

\textbf{2) Company Announcements Corpus (6.4B tokens, 19G)}: Sourced from web-scraped announcements on the official websites of domestic listed companies, this dataset includes a wide array of corporate disclosures from various industries, such as financial reports, significant event statements, notices of shareholders' meetings, stock repurchase plans, and executive changes. It spans 20 years and has been standardized in format to align with the report data format.

\textbf{3) Duxiaoman \citep{zhang2023xuanyuan} Open-Source News FinCorpus (9.6B tokens, 27G)}: Comprising articles and information aggregated from multiple sources, including major financial news websites and social media, this dataset offers a comprehensive collection of financial news and insights. It spans 20 years and has been standardized in format to align with the report data format.

\subsubsection{Filtering Low-quality Fin-Corpus} 
Our 32B financial corpus contains noisy data, including URLs, verbose content, and incoherent or repetitive text. Due to the high cost of using LLMs for full dataset filtering , we distill LLM's ability to judge quality into a lightweight BERT.
We use Qwen2.5-72B-Instruct to rate a 100K-token subset on a 1–10 scale, labeling data above 8 as high-quality and below 4 as low-quality. This yields a 4K-instance training set (2K per class, 90\%–10\% train-test split).
Fine-tuning RoBERTa-wwm-Chinese on this dataset produces a classifier with over 99\% accuracy, which we use to filter the full corpus, removing 15\% of low-quality data.

\subsubsection{Pre-training Details}
Chinese-roberta-wwm-ext \citep{cui2021pre} was used as the initial backbone for FinBERT2 pre-training. Specifically, the training corpus is sliced into the longest contiguous segments of no more than 512 tokens. Strategies \citep{liuRoBERTaRobustlyOptimized2019} such as dynamic masking, pre-training using whole word masking without next sentence prediction (NSP) are also employed, which help minimize training loss and achieve lower bits-per-character (BPC) on the held-out financial corpus. We train with mixed precision and AdamW~\citep{Ilya20270fix} weight decay optimizer on 8 Nvidia A100 40G GPU machines, with the training code developed by HuggingFace's transformer library.

\subsubsection{Expanded Vocabulary for Fin-Tokenizer}
Using the WordPiece algorithm \citep{schuster2012japanese}, we extract domain-specific vocabulary from our 32B-token financial corpus and the ``Colossal, Cleaned, Common Crawl (C4)'' dataset\footnote{\url{https://huggingface.co/datasets/c4}}, the latter being part of the original pre-training data for Chinese-RoBERTa. This process expands the model’s vocabulary by 14,000 words, incorporating a substantial number of high-frequency financial terms and company names (e.g., BYD).
Building on Fin-Tokenizer, we conduct post-pre-training on our financial corpus, yielding FinBERT2, a model better adapted to domain-specific tasks.

\subsection{Task-specific Fine-tuning of Fin-Laberers}

\subsubsection{Five Downstream Datasets for Fin-Labelers}

Due to the disparity between existing public chinese financial datasets and real-world business practices, we constructed five financial classification datasets by directly extracting and annotating data from financial terminal systems. This dataset encompasses three financial application scenarios, including

\textbf{1) Report-related Industry Classification (IC):} Classify report-related passages according to the China International Trust and Investment Corporation (CITIC) Level 1 industry classification, covering 28 industry categories.

\textbf{2) Market Sentiment Classification (MSC):} This task aims to classify the sentiment of textual commentary related to financial events or assets, facilitating market sentiment analysis and stock correlation studies. The first type of sentiment classification, applied to reports, includes four categories that represent varying levels of sentiment polarity and intensity. The second type, focused on news sentiment classification, consists of two categories: positive and negative.

\textbf{3) Named Entity Recognition (NER) in financial reports:} Recognize and extract entities (e.g., company or personal names) appearing in the financial reports.

\begin{table}[tb]
    \centering
    \scriptsize
    \begin{tabular}{lcccccc}
    
    \toprule
        \textbf{Task} & \textbf{Labels} & \textbf{Train/Test Samples} & \textbf{Lr(e-5)} & \textbf{Bs} & \textbf{Epochs} & \textbf{Metric} \\
        \midrule

        \textbf{IC} & 28 & 4000/400 & 5 & 5 & 1 & weighted-f1 \\ 
        \textbf{MSC(4 labels)} & 4 & 1280/400 & 5 & 5 & 1 & weighted-f1 \\ 
        \textbf{MSC(2 labels)} & 2 & 4000/400 & 5 & 5 & 1 & weighted-f1 \\ 
        \textbf{NER(person)} & 3 & 664/140 & 5 & 4 & 5 & recall \\ 
        \textbf{NER(company)} & 3 & 2883/300 & 5 & 4 & 5 & recall \\  
        \bottomrule

    \end{tabular}
    \caption{Details of five classification tasks for Fin-Labelers .}    \label{tab:details1}

\end{table}

\begin{table}[tb!]
\centering
\scriptsize

\begin{tabular}{lcccc} 
\toprule
\textbf{Test Dataset Name}        & \textbf{Query count} & \textbf{Doc count} & \textbf{Avg Query Words} & \textbf{Avg Doc Words}  \\
\midrule
\textbf{Sin-Doc-FinQA}    & 114                     & 1626                 & 29.2                           & 1656.3                        \\
\textbf{Multi-Docs-FinQA} & 54                      & 9384                 & 16.6                           & 589.8                         \\
\textbf{Research Reports}  & 228                     & 1975                 & 28.7                           & 1405.7                        \\
\textbf{Announcements}     & 67                      & 77429                & 40.2                           & 1311.3                        \\
\textbf{Indicators}       & 200                     & 182695               & 27.9                           & 31.8                          \\
\bottomrule
\end{tabular}
\caption{Statistics of FIR-Bench used to evaluate Fin-Retriever.}
\label{tab:details2}
\end{table}

\subsubsection{Fine-tuning Details}
Fine-tuning in BERT usually optimizes all parameters to maximize classification probability. For sequence classification tasks like MSC and IC, the [CLS] token's embedding is fed into a fully connected layer to predict class label probabilities. For token classification tasks like NER, each token's corresponding vector is used for prediction.
In our experiments, we employed a fine-tuning configuration as detailed in \autoref{tab:details1}. 
We fine-tuned 5 Downstream Datasets for fin-Labelers using an AdamW weight decay optimizer on an NVIDIA A100 40GB GPU to ensure consistency and comparability across experiments. For the sequence classification task, we set the following hyperparameters: an epoch of 1, a batch size of 5, and a learning rate of 5e-5. For the token classification task, the same learning rate is maintained, and the epoch was increased to 5.

\subsection{Contrastive Learning for Fin-Retriever}
 We fine-tuned FinBERT2 using contrastive learning on 64,000 financial samples and 150,000 general samples, carefully curated with balanced positives and negatives for optimized performance. The evaluation covers five financial datasets and the C-MTEB benchmark, assessed using recall and nDCG@10.

\subsubsection{Constructing Training Dataset of Fin-Retriever}
To enhance both the financial and general retrieval capabilities of Fin-Retriever, we fine-tuned the pre-trained FinBERT2 on a combination of general-purpose and finance-specific text pairs. The training data comprises 64,000 financial QA data, 100,000 T2Retrieval data, 40,000 MMarcoRetrieval data, and 10,000 DuRetrieval data. Additionally, we optimized the setup and enhanced the quality of positive and negative examples in the training dataset.

    \textbf{1) Mining Sufficient Hard-negatives:}
    Since the collected training datasets lack sufficient negatives, we utilize FinBERT2-base to mine up to 50 hard negatives for each query. Furthermore, financial data is integrated into the sample pool to enhance the dataset's diversity, thereby improving the model's domain-specific capabilities.

    \textbf{2) Balancing Ratio of Positives and Negatives:}
    Achieving a balance between positives and negatives is critical for optimal performance. After extensive experimentation, we found that a ratio of 2 positives to 8 negative instances delivers the best results. Increasing the number of negatives to 15 maintains comparable performance, while reducing it to 5 results in a performance decline. For the Fin-Retriever-large, which has greater capacity than the Fin-Retriever-base, the number of negatives per query is increased from 10 to 15 to better utilize its capacity.

    \textbf{3) Filtering False Positives and False Negatives:}
    In the T2Retrieval, MMarcoRetrieval, and DuRetrieval datasets, positives may not always directly answer the query. To address this issue, we leverage Qwen2.5-72b-Instruct to filter and ensure that positive samples contain sufficient information to effectively respond to the query, thereby enhancing their quality. Similarly, to ensure that negative samples do not include information that could potentially answer the query, we apply further filtering using Qwen2.5-72b-Instruct. By combining high-quality positive and negative samples across datasets, we construct a robust and effective training dataset.

\subsubsection{Contrastive Learning Details}

We train Fin-Retriever based on FinBERT2 using the widely adopted InfoNCE loss \citep{oord2018representation}. We set the following fine-tuning parameters: epoch of 3, batch
size of 512, learning rate of 5e-5, temperature of 0.1, warm-up ratio of 0.1, and weight decay of 0.01. We train the model on 4 NVIDIA A100 40GB GPUs. For long document retrieval, we utilize a sliding window approach, where each window produces an embedding vector per sliding step, with a slight overlap between adjacent windows to maintain semantic coherence. A window size of approximately 400 words and an overlap of 20 words deliver optimal performance.

\subsubsection{Financial Information Retrieval Benchmark (FIR-Bench)}
To comprehensively assess its domain-specific ability, we curated five financial retrieval test datasets (FIR-Bench) derived from our business data, ensuring a thorough evaluation of its performance in the financial domain. The statistics of these five test datasets are presented in \autoref{tab:details2}.

\textbf{1) Single-Document Financial Question-Answer Test Dataset (Sin-Doc FinQA):} This dataset consists of queries linked to both positive and negative document samples. For each query, the candidate documents originate from the same article, with the number of positive documents ranging from a minimum of 1 to a maximum of 10. On average, each query is associated with 8.4 documents, of which 2.6 are positive.

\textbf{2) Multi-Documents Financial Question-Answer Test Dataset (Multi-Docs FinQA):} Unlike Sin-Doc FinQA, the Multi-Docs FinQA has document samples that come from different articles; thus, it contains a much larger number of both positive and negative documents. The maximum number of positive documents is limited to 50. In the corpus, each question is associated with an average of 9,384 documents, of which 14 are labeled as positive on average.

\textbf{3) Financial Retrieval Datasets from Three Sources:} This includes three datasets—financial research reports, indicators, and announcements. We use the cosine similarity metric to calculate the recall rate of positive instances for each dataset.

\subsubsection{General-domain Retrieval Test}

Besides domain-specific retrieval evaluation, we also evaluate the general retrieval capabilities of the models. We utilized a subset of the C-TMEB \citep{xiaoCPackPackagedResources2023} to evaluate model performance in the general domain. Specifically, our models were evaluated on the following representative and influential datasets: T2Retrieval, CovidRetrieval, MMarcoRetrieval, and DuRetrieval datasets. We also used nDCG@10 as the evaluation metric.

\begin{table*}[!h]
\centering
\scriptsize
\resizebox{\textwidth}{!}{
\begin{tabular}{llcccccc}
\toprule
                                        & \textbf{Backbone}                & \multicolumn{1}{l}{\textbf{IC}} & \multicolumn{1}{l}{\textbf{MSC(4 labels)}} & \multicolumn{1}{l}{\textbf{MSC(2 labels)}} & \multicolumn{1}{l}{\textbf{NER(person)}} & \multicolumn{1}{l}{\textbf{NER(company)}} & \multicolumn{1}{l}{\textbf{Avg}}  \\ 
\midrule
\multirow{3}{*}{\textbf{LLMs}}          & \textbf{Qwen2-72b-Instruct}      & 0.9250~                         & 0.4880~                                    & 0.8850~                                    & 0.9669~                                  & \textbf{0.8995~}                          & 0.8329~                           \\
                                        & \textbf{GPT-4-turbo}             & 0.8600~                         & 0.4750~                                    & 0.8880~                                    & 0.9315~                                  & 0.8787~                                   & 0.8066~                           \\
                                        & \textbf{Claude-3.5-Sonnet}       & 0.9030~                         & 0.5230~                                    & 0.8650~                                    & \textbf{0.9957~}                         & 0.8683~                                   & 0.8310~                           \\ 
\midrule
\multirow{3}{*}{\textbf{General BERTs}} & \textbf{BERT-base-chinese}       & 0.9166~                         & 0.8676~                                    & 0.8840~                                    & 0.9901~                                  & 0.8269~                                   & 0.8970~                           \\
                                        & \textbf{Chinese-MacBERT-base}    & 0.9128~                         & 0.8616~                                    & 0.9422~                                    & 0.9854~                                  & 0.8324~                                   & 0.9069~                           \\
                                        & \textbf{Chinese-RoBERTa-wwm-ext} & 0.9196~                         & 0.8841~                                    & 0.9424~                                    & 0.9901~                                  & 0.8158~                                   & 0.9104~                           \\ 
\midrule
\multirow{4}{*}{\textbf{FinBERTs}}      & \textbf{FinBERT1-base}           & 0.9294~                         & 0.9147~                                    & 0.9453~                                    & 0.9901~                                  & 0.8481~                                   & 0.9255~                           \\
                                        & \textbf{Mengzi-BERT-base-fin}    & 0.9083~                         & 0.8657~                                    & 0.9498~                                    & 0.9902~                                  & 0.8324~                                   & 0.9093~                           \\
                                        & \textbf{FinBERT2-base (ours)}    & 0.9398~                          & \textbf{0.9249~}                           & 0.9546~                                    & 0.9901~                                  & 0.8378~                                   & 0.9295~                            \\
                                        & \textbf{FinBERT2–large (ours)}   & \textbf{0.9432~}                & 0.9131~                                    & \textbf{0.9573~}                           & 0.9804~                                  & 0.8514~                                   & \textbf{0.9291~}                  \\
\bottomrule
\end{tabular}
}
\caption{Performance of FinBERT2 and baselines on fin-classification tasks.}
\label{tab:performance1}
\end{table*}

\subsection{Constructing a Pipeline for Fin-TopicModel}

\subsubsection{Overview of Fin-TopicModel}

We developed the Fin-TopicModel on top of Fin-Retriever and other FinBERT2-related components. It performs unsupervised clustering on the Fin-Retriever embeddings using the HDBSCAN algorithm to obtain multiple clusters (topics). For each cluster, it uses c-TF-IDF (Class-Based Term Frequency-Inverse Document Frequency) to measure the importance of words within the cluster. By analyzing the high-frequency vocabulary in each cluster, it automatically generates topic descriptions. 
\subsubsection{Large-scale Unsupervised Title Dataset with Label-free Evaluation}

We constructed a dataset comprising 56,540 report titles extracted from 59,014 articles published between 2022-2024, with an average title length of 27 characters. This dataset was subjected to label-free topic modeling without manual annotation. A comprehensive evaluation is composed of subjective scoring (e.g., coherence, conciseness, and informativeness) using LLMs, clustering metrics like the Silhouette Coefficient and Calinski-Harabasz Index, and additional metrics, including topic diversity and outlier rate, which provide further insights. This framework facilitates a comprehensive and unsupervised exploration of topic modeling in the absence of labeled data.

\subsubsection{Encoding from FinBERT2 Variants}

In conventional practice, retrieval models are often used to extract embeddings from texts to support topic modeling tasks. For this purpose, we selected Fin-Retriever to encode documents. At the same time, we believe that FinLabeler-IC (a fine-tuned industry classification model) allows BERT to further learn semantic information related to specific industry classification tasks during training. These fine-tuned embeddings capture detailed semantic features in identifying the industry, making them highly suitable for topic modeling tasks. Subsequent experiments confirmed that the two chosen models demonstrated diverse advantages in topic modeling.

\subsubsection{Precise Words Cutting}
We used Fin-Labeler-NER to extract 3,290 company-related named entities from titles, which were added to Jieba's custom dictionary to enhance segmentation. To compare Fin-Tokenizer's financial vocabulary with Jieba's default tokenization, we tested Jieba on 13,804 custom financial terms from Fin-Tokenizer, revealing 1,724 tokenization inconsistencies. Analysis showed 545 were two-character terms (mostly auxiliary words or adverbs), and 1,179 were four-character terms (mainly company names).

Fin-Labeler-NER relies on neural methods, while Fin-Tokenizer uses statistical approaches for subword segmentation. To combine the advantages of neural and statistical methods, we designed a merged tokenizer. Using a greedy strategy, it selects token cuts with the longest coverage, thereby enhancing both the accuracy and coverage of NER-based tokenization.

\begin{table*}[!h]
\centering
\scriptsize
\resizebox{\linewidth}{!}{
\begin{tabular}{lrrrcccccccc} 
\toprule
\multirow{2}{*}{\textbf{DR Model}}    & \multicolumn{3}{c}{\textbf{Sin-Doc-FinQA}}          & \multicolumn{2}{c}{\textbf{Multi-Docs-FinQA~ }} & \multicolumn{2}{c}{\textbf{Research Reports}} & \multicolumn{2}{c}{\textbf{\textbf{Announcements}}} & \multicolumn{2}{c}{\textbf{Indicators}}  \\
                                      & \textbf{R@1}    & \textbf{R@3}    & \textbf{R@5}    & \textbf{R@20}   & \textbf{R@50}                 & \textbf{R@10}  & \textbf{R@20}                & \textbf{\textbf{R@10}} & \textbf{R@20}              & \textbf{R@5}    & \textbf{R@10}          \\ 
\midrule
\textbf{BGE-base-zh}                  & 0.479~          & 0.815~          & 0.906~          & 0.238~          & 0.318                         & 0.921          & 0.960~                       & 0.387                  & 0.482~                     & 0.910~          & 0.930~                 \\
\textbf{BCE-embedding-base}           & 0.513~          & 0.824~          & 0.902~          & 0.227~          & 0.309                         & 0.967          & 0.978~                       & 0.318                  & 0.421~                     & 0.803~          & 0.915~                 \\
\textbf{text-embedding-3-small} & 0.511~          & 0.823~          & 0.906~          & 0.197~          & 0.234                         & 0.864          & 0.872~                       & 0.473                  & 0.509~                     & 0.863~          & 0.928~                 \\
\textbf{text-embedding-3-large} & \textbf{0.560~} & 0.845~          & 0.920~          & 0.215~          & 0.257                         & 0.951          & 0.960~                       & 0.492                  & 0.526~                     & 0.940~          & 0.965~                 \\
\textbf{Fin-Retriever-base (Ours)}    & 0.520~          & 0.846~          & 0.916~          & 0.307~          & 0.398                         & \textbf{0.987} & \textbf{0.991~}              & 0.566                  & 0.642~                     & 0.950~          & \textbf{0.975~}        \\
\textbf{Fin-Retriever-large (Ours)}   & 0.554~          & \textbf{0.867~} & \textbf{0.937~} & \textbf{0.315~} & \textbf{0.402}                & 0.983          & 0.987~                       & \textbf{0.571}         & \textbf{0.664~}            & \textbf{0.960~} & 0.970~                 \\
\bottomrule
\end{tabular}
}
\caption{Performance comparison of Fin-Retriever with other dense retrievers on FIR-Bench. We use Recall@k as the evaluation metric.}
\label{tab:performance21}
\end{table*}

\begin{table*}[!h]
\centering
\scriptsize
\resizebox{0.7\linewidth}{!}{
\begin{tabular}{lccccl} 
\toprule
\textbf{DRmodel}                         & \multicolumn{1}{l}{\textbf{T2Retrieval}} & \multicolumn{1}{l}{\textbf{CovidRetrieval}} & \multicolumn{1}{l}{\textbf{MMarcoRetrieval}} & \multicolumn{1}{l}{\textbf{DuRetrieval}} & \textbf{Avg}    \\ 
\midrule
\textbf{BGE-base-zh}               & 0.832                                    & \textbf{0.799}                              & \textbf{0.634}                               & \textbf{0.829}                           & 0.774           \\
\textbf{BCE-embedding-base}     & 0.846                                    & 0.756                                       & 0.634                                        & 0.810                                    & 0.762           \\
\textbf{text-embedding-3-small} & 0.825                                    & 0.706                                       & 0.603                                        & 0.769                                    & 0.726           \\
\textbf{text-embedding-3-large} & \textbf{0.906}                           & 0.760                                       & 0.636                                        & 0.822                                    & \textbf{0.781}  \\
\textbf{Fin-Retriever-base}     & 0.847                                    & 0.779                                       & 0.603                                        & 0.776                                    & 0.751           \\
\textbf{Fin-Retriever-large}    & 0.848                                    & 0.772                                       & 0.603                                        & 0.783                                    & 0.752           \\
\bottomrule
\end{tabular}}
\caption{Performance comparison of Fin-Retriever with other dense retrievers on 4 general retrieval benchmark datasets from C-MTEB. We use NDCG@10 as the evaluation metric.}
\label{tab:performance22}

\end{table*}

\subsubsection{Pipeline Details}

We developed the Fin-TopicModel based on the BERTopic library \citep{grootendorst2022bertopicneuraltopicmodeling}, which is an unsupervised topic modeling library for short texts. First, the embedding model generates semantic vectors for the dataset titles. These embeddings were then reduced to 32 dimensions via UMAP, with parameters set to n\_neighbors=15 and min\_dist=0.0. Next, HDBSCAN was applied for density-based clustering, with min\_cluster\_size=2 and min\_samples=1 to minimize outliers. A custom stopword list based on the \href{https://github.com/CharyHong/Stopwords/blob/main/stopwords_cn.txt}{stopwords\_cn.txt} and the NERcom-Enhanced Tokenizer was used to instantiate a CountVectorizer for text vectorization. This facilitated the use of c-TF-IDF (Class-based Term Frequency-Inverse Document Frequency) to generate a list of keywords for each topic, ranked by their importance, as a descriptive representation of the topics.

\section{Evaluations and Analysis} 

\subsection{Fin-Labelers}
As shown in \autoref{tab:performance1}, we benchmarked FinBERT2 against general-domain Chinese BERTs (e.g., vanilla BERT, MacBERT, and RoBERTa) and financial-domain Chinese BERTs (e.g., FinBERT1 and Mengzi-Fin). 
\autoref{tab:details1} presents a comprehensive summary of the experimental setup, including the number of labels, training and testing sample sizes, fine-tuning hyperparameters (learning rate, batch size, and epochs), and evaluation metrics for each downstream task. A subset of the downstream task datasets was held out for evaluating the Fin-Labeler models. Model performance was assessed throughout the training process, with the optimal test results being recorded. All BERT-based models underwent identical fine-tuning procedures.

We also benchmark leading LLMs, including Qwen2-72b-Instruct, GPT-4-turbo, and Claude-3.5-Sonnet. For each task, we meticulously crafted prompts, employing popular techniques such as role specification, few-shot learning, and chain-of-thought to optimize LLM performance. To ensure the fairness and accuracy of the evaluation, we employed identical prompts across three robust LLMs and had each model assess every sample three times. The final predicted label was determined by applying a majority voting rule to the obtained labels. The prompts are shown in \autoref{prompts:classi}.

\subsubsection{Compared with LLMs}

As shown in \autoref{tab:performance1}, the average F1 scores of Qwen2-72b-Instruct (0.8329), GPT-4-turbo (0.8066), and Claude-3.5-Sonnet (0.8310) are significantly lower than FinBERT2-base (0.9295) and FinBERT2-large (0.9291). On the challenging market sentiment classification task, LLMs scored below 0.523, far behind FinBERT2-base (0.9249). For NER tasks, LLMs showed superiority—e.g., Claude-3.5-Sonnet excelled in company names (0.868).
These findings highlight the limitations of LLMs on domain-specific tasks without fine-tuning and reinforce FinBERT2's superiority in both effectiveness and efficiency for financial applications.

\subsubsection{Compared with Other BERTs}
General BERTs achieve average scores of 0.8970–0.9104 across five tasks, while FinBERT2-base and FinBERT2-large outperform them with scores of 0.9295 and 0.9291. For the complex four-class market sentiment classification (MSC) task, FinBERT2-base achieves 0.9249, compared to 0.8841 from general-purpose models, highlighting its ability to capture domain-specific nuances.
Compared to other FinBERTs like FinBERT1-base and Mengzi-BERT-base-fin, FinBERT2 also performs better. For example, in the IC task, FinBERT2-base scores 0.9398, surpassing FinBERT-Chinese (0.9294) and Mengzi-BERT-base-fin (0.9083).
While simpler tasks show comparable performance, FinBERT2 demonstrates clear superiority in complex tasks such as four-class MSC, underscoring its domain-specific advantages.

\subsubsection{Analysis about Different Tasks and Architecture}
 In the context of simple NER tasks, such as person name recognition, some LLMs may exhibit entity omission issues, resulting in performance inferior to that of fine-tuned BERT models. However, in complex NER tasks, such as company name recognition, LLMs are better able to demonstrate stronger generalization capabilities. This indicates that the performance of LLMs and BERT-based models varies significantly depending on the complexity of the NER task. Furthermore, FinBERT2-large outperforms FinBERT2-base on complex NER tasks. This indicates that increased model scale contributes to improved generalization for complex NER tasks.

MSC(4 labels) presents a distinct challenge for LLMs due to its reliance on industry-specific annotation criteria. The annotation criteria comes from professional analysts and exhibit certain industry-specific characteristics. For instance, in traditional industries, 10\% growth might be considered highly positive, whereas in another industry, it might only be viewed as moderately positive. The in-context learning capabilities of LLMs are frequently confined to surface-level semantics and numerical values. This limitation restricts their adaptability to tasks that involve nuanced, industry-specific criteria, which are often challenging to express through prompts.

\begin{table*}[!h]
\centering
\scriptsize
\resizebox{\linewidth}{!}{
\begin{tabular}{lccc|ccc} 
\toprule
\multirow{2}{*}{\textbf{Embedder in pipeline}} & \multicolumn{3}{c}{\textbf{Avg LLM-score (0-3)}}                        & \multicolumn{3}{c}{\textbf{Cluster Qulity Metrics}}                                                 \\
                                            & \textbf{Coherence} & \textbf{Conciseness} & \textbf{Informativity} & \textbf{Silhouette Coefficient} & \textbf{Calinski Harabasz Score} & \textbf{Davies Bouldin Score}  \\
\midrule
\textbf{BGE-base-zh}                           & 1.765~             & 1.550~               & 1.865~                 & 0.141~                          & 11.394~                          & 1.254~                         \\
\textbf{BCE-embedding-base}                           & 1.790~             & 1.460~               & 1.870~                 & 0.171~                          & 12.483~                          & 1.158~                         \\
\textbf{text-embedding-3-small}      & 1.744~             & 1.533~               & 1.809~                 & 0.106~                          & 11.204~                          & 1.274~                         \\
\textbf{text-embedding-3-large}      & 1.795~             & 1.445~               & 1.825~                 & 0.182~                          & 12.934~                          & 1.155~                         \\
\textbf{Fin-Labeler-IC}                     & 1.830~             & 1.500~               & 1.845~                 & 0.170~                          & 11.108~                          & 1.108~                         \\
\textbf{Fin-Retriever-base}                 & \textbf{1.835~}    & 1.515~               & \textbf{1.905~}        & \textbf{0.192~}                 & \textbf{13.296~}                 & 1.077~                         \\
\textbf{Fin-Retriever-large}                & 1.760~             & \textbf{1.570~}      & 1.860~                 & 0.174~                          & 12.690~                          & \textbf{1.070~}                \\
\bottomrule
\end{tabular}
}
\caption{Performance comparison of different embedding models on the topic modeling task. The LLM-scores reflect the semantic quality of the topics, while the clustering quality metrics measure the compactness and separation of the topics.}
\label{tab:performance31}

\end{table*}

\begin{table}[!h]
\centering
\scriptsize
\resizebox{\linewidth}{!}{
\begin{tabular}{lccccc} 
\toprule
\textbf{Embedder in pipeline}             & \textbf{TD} & \textbf{Outliers Rate} & \textbf{Topic Count} & \textbf{Avg Docs/ Topic} & \textbf{SD Docs/Topic}  \\
\midrule
\textbf{BGE-base-zh}                      & 0.211~                   & 0.230~                 & 11872                   & 3.563~                                             & 3.125~                                \\
\textbf{BCE-embedding-base}                      & 0.217~                   & 0.253~                 & 12129                   & 3.667~                                             & 3.281~                                \\
\textbf{text-embedding-3-small}& 0.209~                   & 0.246~                 & 12260                   & 3.477~                                             & 2.962~                                \\
\textbf{text-embedding-3-large}& \textbf{0.218~}          & 0.224~                 & 11820                   & 3.712~                                             & 3.353~                                \\
\textbf{Fin-labeler-IC}                & 0.196~                   & \textbf{0.222~}        & \textbf{13417}          & \textbf{3.280~}                                    & \textbf{2.144~}                       \\
\textbf{Fin-Retriever-base}            & \textbf{0.218~}          & \textbf{0.222~}        & 11796                   & 3.728~                                             & 3.211~                                \\
\textbf{Fin-Retriever-large}           & 0.213~                   & 0.227~                 & 12008                   & 3.637~                                             & 3.178~                                \\
\bottomrule
\end{tabular}
}
\caption{Summarization of the statistical properties of the topics generated in the topic modeling task. }
\label{tab:performance32}
\end{table}

\subsection{Fin-Retriever}
We compare our Fin-Retriever with popular open-source models of similar parameter sizes (BAAI General Embedding (BGE), Bidirectional Contrastive Embedding (BCE)) as well as OpenAI's proprietary models \footnote{https://platform.openai.com/docs/guides/embeddings}.

\subsubsection{Performance on FIR-Bench}
As shown in \autoref{tab:performance21}, both Fin-Retriever-base and Fin-Retriever-large outperform general-purpose retrievers in FIR-Bench, demonstrating their strong domain specialization. Fin-Retriever-large consistently achieves the highest recall scores, with an average R@k of 0.746, while Fin-Retriever-base also surpasses most baselines, achieving 0.730 on average. Compared to the best-performing general model, text-embedding-ada-002-large (0.723 avg.), both Fin-Retriever variants exhibit superior financial retrieval capabilities. Particularly in Research Reports (R@10 = 0.987 for Fin-Retriever-large vs. 0.951 for text-embedding-ada-002-large) and Announcements (R@10 = 0.571 for Fin-Retriever-large vs. 0.492 for text-embedding-ada-002-large), the Fin-Retriever models demonstrate a significant advantage. This suggests that in financial domains where precision is critical, Fin-Retriever provides a substantial improvement over general-purpose dense retrievers.

\subsubsection{Performance on General Retrieval Datasets}
Despite being optimized for financial retrieval, Fin-Retriever-base and Fin-Retriever-large remain competitive in general retrieval tasks, as reflected in \autoref{tab:performance22}. Fin-Retriever-large achieves an average NDCG@10 score of 0.752, while Fin-Retriever-base follows closely with 0.751, both outperforming text-embedding-3-small (0.726) and approaching the performance of text-embedding-3-large (0.781), an advanced model from OpenAI’s text-embedding series, with a parameter count significantly exceeding that of BERT.

\subsection{Fin-TopicModel}

\subsubsection{Metrics}

Evaluating topic models is a complex and evolving challenge. We introduce a suite of unsupervised evaluation metrics for Fin-TopicModel, which is shown in \autoref{tab:performance31} and \autoref{tab:performance32}, thus circumventing the need for labeled data. The metrics fall into three broad categories. First, subjective scoring leverages LLMs to evaluate topic descriptions such as coherence, conciseness, and informativeness. Second, the clustering quality metrics, including the Silhouette Coefficient and Calinski-Harabasz Index, are employed. Third, supplementary metrics like topic diversity and outlier rate are incorporated to provide a more nuanced understanding. For a comprehensive description of the metrics, refer to \autoref{metric}. 

\subsubsection{Analysis of Fin-TopicModel based on Fin-Retrievers}
As shown in \autoref{tab:performance31} and \autoref{tab:performance32}, the Fin-TopicModel based on Fin-Retriever-base demonstrates all-around superior performance across both subjective scoring (coherence, conciseness, and informativeness) and clustering quality metrics. It achieves the highest coherence score (1.835) and excels in clustering compactness and separation, as reflected by the highest Calinski-Harabasz Index (13.296) and a low Davies-Bouldin Score (1.077). These results highlight the better capabilities of Fin-Retriever-base in topic modeling tasks. In addition, it achieves a high topic diversity score (0.218) while maintaining the lowest outlier rate (0.222). By contrast, Fin-Retriever-large also performs well but falls slightly short of Fin-Retriever-base. This performance gap may be attributed to the increased window length, resulting in a disadvantage for short text representation. 

\subsubsection{Analysis of TopicModel based on Fin-Labeler-IC}

The Fin-TopicModel based on Fin-Labeler-IC showcases unique strengths and trade-offs in topic modeling. It generates the highest number of topics (13,417), significantly more than other models, which suggests its potential to capture granular distinctions in the data. 
Additionally, it achieves competitive semantic quality and clustering quality, with coherence (1.830) and informativeness (1.845) scores. However, this granularity comes at the cost of lower topic diversity (0.196), indicating potential redundancy or over-segmentation of the topics. Despite its simpler fine-tuning approach, Fin-Labeler-IC is competent for topic modeling by leveraging its industry-specific training objective, which is inherently aligned with topic relevance. This trade-off highlights the value of task-specific embeddings for applications.

\subsection{Ablations and Discussions}

We conducted a detailed ablation study to verify the impact of pre-training corpus size and vocabulary expansion on the final results. All ablation experiment results are presented in \autoref{tabs:ablation}.

\subsubsection{Ablations of Domain Pre-trained Data Volume}

To demonstrate that increasing the volume of the domain pre-training corpus can enhance model capability, we trained the 7.3B/10.3B FinBERT models with the same configuration on a subset of the corpus. We then evaluate these models on both classification and retrieval tasks.

\autoref{abalation1} demonstrates the impact of pre-training data volume on classification tasks. The results indicate a clear performance improvement as the volume of the pre-training corpus increases. Without domain-specific pre-training, the average F1 score is 0.9104. With 7.3B tokens, the score increases to 0.9189, and with 32B tokens (FinBERT2-base), it reaches 0.9295, achieving the best results. Tasks like MSC and IC benefit significantly from larger pre-training datasets, showing notable performance gains. This highlights the crucial role of extensive pre-training in capturing domain-specific features and improving generalization in classification tasks.

\autoref{abalation2} presents the effect of pre-training data volume on retrieval tasks, evaluated using Recall@k. The results show that larger pre-training volume lead to substantial performance improvements. Without domain-specific pre-training, the average recall is 0.621. Pre-training on 16B and 32B tokens improves the average recall to 0.659 and 0.686, respectively. Tasks such as 'Announcements' and 'Research Reports' benefit the most, with recall@10 increasing from 0.325 to 0.566 and 0.943 to 0.987, respectively. These results demonstrate that larger pre-training volume enable the model to understand financial texts better and enhance retrieval performance.

\subsubsection{Ablation of Vocabulary Expansion on Retrieval Performance}

\autoref{abalation4} presents a comparison of pre-training FinBERT2-base with and without vocabulary expansion. The results show that adding a domain-specific vocabulary (Fin-Tokenizer) significantly improves performance across retrieval tasks. For instance, the model with Fin-Tokenizer achieves R@50 of 0.397 on Multi-Docs-FinQA and R@20 of 0.642 on Announcements, representing increases of 0.034 and 0.155 compared to the model without Fin-Tokenizer. This highlights that vocabulary expansion effectively enhances the model's ability to understand financial terminology. 

\subsubsection{Is Advantage in Retrieval Tasks Attributable to Fine-Tuning Rather than the FinBERT2 Backbone?}

Although our model outperforms others on multiple financial retrieval tasks, this advantage may be partly attributed to the lack of fine-tuning of other models on the same financial data. To ensure a fair comparison, we fine-tuned the BGE-base-zh using the same dataset. As shown in \autoref{abalation3}, BGE-base-zh, which shares the same architecture as our Fin-Retriever-base, was fine-tuned with the same data and hyperparameters employed in the second stage of contrastive learning for Fin-Retriever. We subsequently evaluated both models on FIR-Bench. These results suggest that FinBERT2-base is well-suited for retrieval tasks, especially for task of research reports and announcements.

\subsubsection{Is BERT-large a Better Backbone Than BERT-base?}

We conducted the whole pipeline based on both the RoBERTa-chinese-base and RoBERTa-chinese-large respectively, and systematically compared their performance. While the results demonstrate that the large one generally outperforms its base counterpart in both classification tasks and retrieval tasks, the performance gains are not significant. Furthermore, training and fine-tuning the large model require extensive hyperparameter optimization and computational resources, posing a considerable challenge. Due to these factors, we did not dedicate substantial effort to further optimizing the large model. Despite the slight advantage shown by the large variant, the base model remains highly competitive and efficient, making it a strong choice for practical applications.

\section{Limitations and Future Work}

The study has several limitations, which are outlined below:

1) Domain Specificity: The model is primarily trained on financial text data, so its performance on non-financial tasks could be suboptimal compared to general-purpose models.

2) Performance Variability in Downstream Tasks: Although fine-tuning enhances the model’s accuracy in financial text classification, the results still exhibit a degree of randomness, suggesting potential instability in certain applications.

3) Challenges in Topic Modeling: The use of HDBSCAN clustering has issues such as topic redundancy or over-segmentation.

The future-oriented works are outlined below:

1) Pre-Training Optimization: We plan to enhance domain adaptation by leveraging advanced BERT-based architectures, such as ModernBERT\citep{warner2024smarter}, or BERT with larger-scale such as MegatronBert-1.3B\citep{Fengshenbang-LM}.

2) Better Fin-retriever: The dataset trained in the Fin-retriever can be collected more comprehensively to further improve the performance.

3) Topic Modeling Refinement: Additional optimization techniques remain unexplored, such as outlier reduction strategies, mitigation of topic redundancy or over-segmentation, and advanced topic description.

4) More Applications: Beyond topic modeling, we aim to deploy FinBERT2 in quantitative finance. For instance, Fin-Labeler-MSC, fine-tuned on market sentiment data, exhibits strong financial sentiment understanding and could serve as a predictive factor for selecting high-performing stocks with excess returns.

\section{Conclusion}

In this work, we introduced FinBERT2, a specialized bidirectional encoder pre-trained on the largest known Chinese financial corpus (32B tokens). Our results demonstrate that encoder-only models still play a crucial role in financial NLP, complementing the strengths of decoder-only LLMs. Specifically, FinBERT2 achieves (1) superior performance on discriminative financial classification tasks, outperforming existing (Fin)BERT variants and leading LLMs; (2) enhanced retrieval capabilities through Fin-Retrievers, surpassing both open-source and proprietary embedding models; and (3) improved topic modeling with Fin-TopicModel, yielding better clustering and topic representation. These findings highlight the continued relevance of encoder-based architectures in financial AI, particularly in scenarios requiring high precision and domain-specific understanding. 

\begin{acks}

This work was supported in part by the National Key Research and Development Program of China under Grant 2022YFC3303301, Grant 2023YFC3305401, and Grant 2023YFC3305402 and in part by the National Natural Science Foundation of China (Nos. 62302059 and 62172053).

\end{acks}

\clearpage

\bibliographystyle{ACM-Reference-Format}
\balance
\bibliography{custom}

\appendix



\section{Prompts of Downstream Tasks using LLMs(Chinese to English
already)}
\label{prompts:classi}
\subsection{IC Prompt}
\textbf{Role:} Senior Industry Researcher.  
\textbf{Task:} According to the CITIC industry classification definition, classify short financial texts into primary CITIC industry categories.  

The industry list and classification descriptions are as follows:  
Building Materials: Involves the production and sale of construction materials such as cement, glass, ceramics, and new building materials.  
Food and Beverages: Includes food manufacturing and beverage production, such as liquor, dairy products, and meat processing.  
…… (More than 20 industries are omitted)

\textbf{Input:}  
text: Financial news brief  

\textbf{Result Format:}  
Rationale:\{CoT, no more than 100 words\} 
Result:\{Industry Name\}

\subsection{MSC(2 labels) Prompt}
\textbf{Role:} Senior Financial Data Analyst, Senior Industry Researcher.  
\textbf{Task:} Financial sentiment classification: Aim to classify evaluative texts on financial events or items into sentiments to observe market sentiment.  

The sentiment classification task includes two categories:  
0: Negative  
1: Positive  

\textbf{Result Format:}  
Rationale:\{CoT, no more than 100 words\}  
Result:\{Sentiment label, 0 or 1\}

\subsection{MSC(4 labels) Prompt}
\textbf{Role:} Senior Financial Data Analyst, Senior Industry Researcher.  
\textbf{Task:} Financial sentiment classification: Aim to classify evaluative texts on financial events or items into sentiments to observe market sentiment.  

The sentiment classification task includes four categories:  
Positive Sentiment (Label: 3): Texts typically include positive evaluations of company performance, stock recommendations, or optimistic industry outlooks. Examples include terms such as "recommend," "exceeds expectations," "significant growth," "strongly recommend," "high-quality," "leader," and "performance surge," indicating positive evaluations and optimistic expectations of companies or industries.  
Neutral-Positive Sentiment (Label: 2): Texts may include affirmations of company performance but also concerns or uncertainties about certain factors. Such texts often use terms like "in line with expectations," "stable," "neutral-positive," "slight growth," and "maintain," reflecting confidence in companies or industries but with some reservations.  
Neutral Sentiment (Label: 1): Texts provide objective descriptions of companies or industries without obvious positive or negative sentiment tendencies. Terms like "neutral," "stable," "flat," and "basically in line with expectations" indicate neutral views on companies or industries.  
Negative Sentiment (Label: 0): Texts typically contain concerns about company performance, non-recommendations of stocks, or pessimistic industry outlooks. Examples include terms like "decline," "losses," "risks," "below expectations," "reduction," and "negative growth," reflecting negative evaluations and pessimistic expectations of companies or industries.  

\textbf{Result Format:}  
Rationale:\{CoT, no more than 100 words\}
Result:\{Sentiment label, one of 0-3\}

\subsection{NER(company) Prompt}
\textbf{Role:} Senior Data Annotation Engineer.  
\textbf{Task:} NER (Named Entity Recognition): Identify and extract company entities mentioned in financial texts.

\textbf{Example:}……

\textbf{Result Format:}  
Rationale:\{CoT, no more than 100 words\}  
Result:\{List of company entities\}

\subsection{NER(person) Prompt}
\textbf{Role:} Senior Data Annotation Engineer, Entity Extraction Engineer.  
\textbf{Task:} NER (Named Entity Recognition): Identify and extract name entities mentioned in financial texts.  

\textbf{Example:}……

\textbf{Result Format:}  
Rationale:\{CoT, no more than 100 words\}  
Result:\{List of name entities\}
\textbf{Example:}……

\section{Metrics of Fin-TopicModel}
\label{metric}
\begin{itemize}
    \item \textbf{Subjective Evaluation:} We leverage a Qwen-max to subjectively evaluate the topic descriptor lists. First, 200 topic descriptor lists are randomly sampled, and Qwen-max is prompted to score them on three aspects: \textbf{Coherence}, \textbf{Conciseness}, and \textbf{Informativity} (on a scale of 1 to 3). The average score across these three dimensions is calculated and used as the final metric.
    
    \item \textbf{Clustering Quality Evaluation:} We adopt three widely used metrics—\textbf{Silhouette Coefficient}, \textbf{Calinski-Harabasz Index}, and \textbf{Davies-Bouldin Index}—to evaluate the quality of the embeddings used for clustering. These metrics measure the compactness of data within clusters and the separation between clusters, providing a quantitative assessment of the clustering algorithm's effectiveness.
    
    \item \textbf{Topic Diversity (TD):} This metric calculates the proportion of unique words across topics. TD values range from 0 to 1, where higher values indicate greater topic diversity and more varied topics generated by the model.

    \item \textbf{Outlier Rate:} We also calculate the proportion of outliers (documents assigned a label of -1) relative to the total dataset size as an additional measure of embedding quality. By the way, the number of outliers can be reduced through post-prediction strategies or parameter adjustments.
    
    \item \textbf{Other statistics:} In addition, we summarize other statistical properties of the topics generated by different embedding models in the topic modeling task. These include the \textbf{outlier rate}, \textbf{total number of topics}, \textbf{average document count per topic}, and the \textbf{standard deviation of document counts per topic}, which provide further insights into the characteristics of the generated topics.
\end{itemize}
\section{Ablations tables}
\label{tabs:ablation}

\begin{table*}[!h]
\centering
\resizebox{\linewidth}{!}{
\begin{tabular}{lcccccc} 
\toprule
\textbf{Backbone for task-specific fine-tuning}               & \textbf{IC}      & \textbf{MSC(4 labels)} & \textbf{MSC(2 labels)} & \textbf{~NER(person)~} & \textbf{~NER(company)~} & \textbf{Avg}     \\ 
\midrule
\textbf{w/o domain-pre-trained (Chinese-RoBERTa)}              & 0.9196~          & 0.8841~                & 0.9424~                & 0.9901~                & 0.8158~                 & 0.9104~          \\
\textbf{w/ 7.3B domain-pre-trained}                             & 0.9241~          & 0.9044~                & 0.9424~                & \textbf{0.9902~}       & 0.8333~                 & 0.9189~          \\
\textbf{w/ 10.3B domain-pre-trained}                            & 0.9252~          & 0.8975~                & 0.9499~                & \textbf{0.9902~}       & 0.8278~                 & 0.9181~          \\
\textbf{w/ 16B domain-pre-trained}                              & \textbf{0.9437~} & 0.9176~                & 0.9525~                & \textbf{0.9902~}       & 0.8333~                 & 0.9275~          \\
\textbf{\textbf{w/ 32B domain-pre-trained (our FinBERT2-base)}} & 0.9398~          & \textbf{0.9249~}       & \textbf{0.9546~}       & 0.9901~                & \textbf{0.8378~}        & \textbf{0.9295~}  \\
\bottomrule
\end{tabular}
}
\caption{Ablation experiments about pre-training strategy and data volume on different backbones following the same task-specific fine-tuning procedure. w * means with *,w/o * means without *.    }
\label{abalation1}
\end{table*}

\begin{table*}[!h]
\centering
\resizebox{\linewidth}{!}{
\begin{tabular}{lcccccccccc} 
\toprule
\multirow{2}{*}{\textbf{Backbone for contrast fine-tuning}} & \multicolumn{3}{c}{\textbf{Sin-Doc-FinQA}}       & \multicolumn{2}{c}{\textbf{Multi-Docs-FinQA}} & \multicolumn{2}{c}{\textbf{Research Reports}} & \multicolumn{2}{c}{\textbf{Announcements}}      & \multirow{2}{*}{\textbf{Avg}}  \\
                                                            & \textbf{R@1}   & \textbf{R@3}   & \textbf{R@5}   & \textbf{R@20}  & \textbf{\textbf{R@50}}       & \textbf{R@10}  & \textbf{\textbf{R@20}}       & \textbf{\textbf{R@10}} & \textbf{\textbf{R@20}} &                                \\
\midrule
\textbf{w/o domain-pre-trained (Chinese-RoBERTa)}           & \textbf{0.534} & 0.837          & 0.911          & 0.291          & 0.376                        & 0.943          & 0.956                        & 0.325                  & 0.417                  & 0.621~                         \\
\textbf{w/ 7.3B domain-pre-trained}                         & 0.507          & 0.838          & 0.915          & 0.292& 0.384& 0.934          & 0.947                        & 0.412                  & 0.445                  & 0.630~\\
\textbf{w/ 10.3B domain-pre-trained}                        & 0.522          & 0.819          & \textbf{0.920} & 0.296& 0.388& 0.965          & 0.969                        & 0.390                  & 0.448                  & 0.635~\\
\textbf{\textbf{w/ 16B domain-pre-trained}}                 & 0.514          & 0.844          & 0.916          & 0.305          & 0.392                        & 0.965          & 0.982                        & 0.462                  & 0.549                  & 0.659~                         \\
\textbf{w/ 32B domain-pre-trained (our FinBERT2-base)}      & 0.520          & \textbf{0.846} & 0.916          & \textbf{0.307} & \textbf{0.398}               & \textbf{0.987} & \textbf{\textbf{0.991}}      & \textbf{0.566}         & \textbf{0.642}         & \textbf{0.686}                \\
\bottomrule
\end{tabular}
 }
\caption{Ablation experiments about data volume of domain-pre-train on different backbones after contrast fine-tuning. w * means with *,w/o * means without *. We use Recall@k as the metric.}
\label{abalation2}
\end{table*}

\begin{table*}[!h]
\centering
\resizebox{0.9\linewidth}{!}{
\begin{tabular}{lccccccccc} 
\toprule
\multirow{2}{*}{\textbf{FinBERT2-base~pre-trained Configuration}} & \multicolumn{3}{c}{\textbf{Sin-Doc-FinQA}}         & \multicolumn{2}{c}{\textbf{Multi-Docs-FinQA~ }}            & \multicolumn{2}{c}{\textbf{Research Reports}} & \multicolumn{2}{c}{\textbf{Announcements}}                                                                               \\ 
\cmidrule{2-10}
                                                                 & \textbf{R@1}   & \textbf{R@3}    & \textbf{R@5}    & \textbf{~R@20}                            & \textbf{~R@50} & \textbf{~R@10} & \textbf{~R@20}               & \textbf{\textbf{\textbf{\textbf{R@10}}}} & \textbf{\textbf{\textbf{\textbf{\textbf{\textbf{\textbf{\textbf{R@20}}}}}}}}  \\ 
\midrule
\textbf{w/o vocabulary expansion}                               & \textbf{0.521} & 0.843~          & \textbf{0.917~} & \textbf{\textbf{\textbf{\textbf{0.279}}}} & 0.363          & 0.896          & 0.938                        & 0.408                                    & 0.487                                                                         \\
\textbf{w/ vocabulary expansion (Fin-Tokenizer)}                & 0.520          & \textbf{0.846~} & 0.916~          & 0.307                                     & \textbf{0.398} & \textbf{0.987} & \textbf{0.991}               & \textbf{0.566}                           & \textbf{\textbf{0.642}}                                                       \\
\bottomrule
\end{tabular}
}
\caption{Ablation experiments about tokenizer w/wo vocabulary expansion.}
\label{abalation4}

\end{table*}

\begin{table*}[!h]
\centering
\resizebox{0.9\linewidth}{!}{
\begin{tabular}{lccccccccc} 
\toprule
\multirow{2}{*}{\textbf{Backbone for contrast fine-tuning}} & \multicolumn{3}{c}{\textbf{Sin-Doc-FinQA}}          & \multicolumn{2}{c}{\textbf{Multi-Docs-FinQA~ }}            & \multicolumn{2}{c}{\textbf{Research Reports}} & \multicolumn{2}{c}{\textbf{Announcements}}                                                                               \\ 
\cmidrule{2-10}
                                                            & \textbf{R@1}    & \textbf{R@3}    & \textbf{R@5}    & \textbf{~R@20}                            & \textbf{~R@50} & \textbf{~R@10} & \textbf{~R@20}               & \textbf{\textbf{\textbf{\textbf{R@10}}}} & \textbf{\textbf{\textbf{\textbf{\textbf{\textbf{\textbf{\textbf{R@20}}}}}}}}  \\ 
\midrule
\textbf{BGE-base-zh}                                        & 0.519~          & 0.843~          & \textbf{0.920~} & \textbf{\textbf{\textbf{\textbf{0.320}}}} & 0.389          & 0.965          & 0.978                        & 0.378                                    & 0.442                                                                         \\
\textbf{FinBERT2-base}                                      & \textbf{0.520~} & \textbf{0.846~} & 0.916~          & 0.307                                     & \textbf{0.398} & \textbf{0.987} & \textbf{0.991}               & \textbf{0.566}                           & \textbf{\textbf{0.642}}                                                       \\
\bottomrule
\end{tabular}
}
\caption{Performance of BGE-base-zh and FinBERT2-base after fin-retrieval fine-tuned by the same training dataset and fine-tuning processes. We use Recall@k as the metric.}
\label{abalation3}
\end{table*}

\newpage

\section{Prompts for Assessing Topic Descriptive Words (Translated from Chinese)}
Please evaluate the given topic keyword list based on the following standards for topic quality assessment. For each criterion, provide a score ranging from 1 to 3, along with a brief explanation of the score.

\textbf{Topic Quality Assessment Criteria:}  
1. Coherence
   Definition: The keywords within a topic should be semantically related and collectively describe a topic or multiple closely related topics.
   
2. Conciseness
   Definition: A topic should not contain irrelevant or meaningless words, such as noise words or semantically redundant terms.
   
3. Informativity
   Definition: A topic should provide sufficient, specific, meaningful, or valuable information, covering different aspects of the same topic.
   
\textbf{Evaluation Instructions:}  
For the provided topic keyword list, rate each criterion on a scale of 1 to 3:
1 point: Poor performance, does not meet the standard.
2 points: Average performance, partially meets the standard.
3 points: Excellent performance, fully meets the standard.
For each rating, provide a brief explanation to justify the score.
Input:\{Topic Keywords List\}

\textbf{Example Response Format:}  
\{
  "Topic Keyword List": ["strategy", "market", "investment", "risk", "return"],
  "Evaluation": \{
    
    "Coherence": \{
      "Score": 3,
      "Explanation": "Keywords are closely related, all relevant to the field of financial investment, and collectively describe the theme of investment strategies."
    \},
    
    "Conciseness": \{
      "Score": 3,
      "Explanation": "Keywords are clear, with no stopwords or meaningless terms, and no redundancy detected."
    \},
    
    "Informativity": \{
      "Score": 2,
      "Explanation": "The topic only covers the main aspects of financial investment but lacks detailed descriptions of specific markets or investment tools."
    \}
  \}
\}

\end{document}